\documentclass[prd,aps,lengthcheck,twocolumn,nofootinbib,notitlepage,floatfix,superscriptaddress]{revtex4-2}
\usepackage{graphics,graphicx}
\usepackage{amsmath, amssymb}
\usepackage{multirow}
\usepackage{color}
\usepackage{verbatim}
\usepackage{hyperref}
\usepackage[normalem]{ulem}  
\usepackage{ulem}
\usepackage{color}
\usepackage{cancel}
\usepackage{mathtools}

\renewcommand\sout{\bgroup\color{blue} \ULdepth=-.5ex \ULset}
\def\slashchar#1{\setbox0=\hbox{$#1$}  
\dimen0=\wd0     
\setbox1=\hbox{/} \dimen1=\wd1  
\ifdim\dimen0>\dimen1   
\rlap{\hbox to \dimen0{\hfil/\hfil}} 
#1     
\else     
\rlap{\hbox to \dimen1{\hfil$#1$\hfil}} 
/      
\fi}

\newcommand{\dd}{\mathrm{d}}

\begin{document}

\title{Probing nuclear liquid-gas phase transition with isospin correlations}

\date{\today}
\author{Micha\l{} Marczenko}
\email{michal.marczenko@uwr.edu.pl}
\affiliation{Incubator of Scientific Excellence - Centre for Simulations of Superdense Fluids, University of Wroc\l{}aw, plac Maksa Borna 9, PL-50204 Wroc\l{}aw, Poland}
\author{Krzysztof Redlich}
\affiliation{Institute of Theoretical Physics, University of Wroc\l{}aw, plac Maksa Borna 9, PL-50204 Wroc\l{}aw, Poland}
\affiliation{Polish Academy of Sciences PAN, Podwale 75, 
PL-50449 Wroc\l{}aw, Poland}
\author{Chihiro Sasaki}
\affiliation{Institute of Theoretical Physics, University of Wroc\l{}aw, plac Maksa Borna 9, PL-50204 Wroc\l{}aw, Poland}
\affiliation{International Institute for Sustainability with Knotted Chiral Meta Matter (WPI-SKCM$^2$), Hiroshima University, Higashi-Hiroshima, Hiroshima 739-8526, Japan}

\begin{abstract}
We investigate the fluctuations of the net-baryon number near the critical point of the liquid-gas phase transition. We use the parity doublet model in the mean-field approximation fixed to the zero-temperature properties of nuclear matter to account for critical behavior. We explicitly calculate the fluctuations of the net-proton and net-neutron numbers as well as their correlations in the isospin-symmetric matter. We focus on the qualitative properties and systematics of the first- to fourth-order susceptibilities and their ratios. We demonstrate that the fluctuations of net-proton number do not reflect the total net-baryon number fluctuations in the vicinity of the nuclear liquid-gas phase transition. We also study the behavior of the baryon and proton number factorial cumulants. We highlight the importance of the non-trivial correlations between protons and neutrons.
\end{abstract}
\maketitle

\section{Introduction}
\label{sec:intro}

One of the goals of high-energy physics is to determine the phase diagram of quantum chromodynamics (QCD)~(see, e.g.,~\cite{Du:2024wjm} for a recent theoretical overview). Studies of QCD on the lattice (LQCD) were crucial in determining that at finite temperature and vanishing chemical potential, the transition from hadronic matter to quark-gluon plasma is a smooth crossover~\cite{Bazavov:2014pvz, Borsanyi:2018grb, Bazavov:2017dus, Bazavov:2020bjn, Bazavov:2020bjn, Aoki:2006we}. At finite chemical potential, the phase structure is potentially more complex. Effective models~\cite{Bowman:2008kc, Ferroni:2010ct, Klevansky:1992qe, Buballa:2003qv, Fukushima:2003fw, Ratti:2005jh, Skokov:2010uh}, Dyson-Schwinger approach~\cite{Qin:2010nq, Gao:2015kea, Fischer:2014ata}, and the Functional Renormalization Group~\cite{Shi:2014zpa} suggest the existence of a first-order QCD phase transition at finite chemical potential, and therefore the existence of a critical point in the QCD phase diagram. 

Locating a putative QCD critical point is part of the experimental effort in large-scale heavy-ion collision experiments. It was suggested that the phase structure of QCD can be investigated through fluctuations and correlations of conserved charges. They are known to be promising theoretical observables in search of critical behavior at the QCD phase boundary~\cite{Stephanov:1999zu, Asakawa:2000wh, Hatta:2003wn, Friman:2011pf}. In particular, fluctuations of conserved charges have been proposed to probe the QCD critical point, as well as the remnants of the $O(4)$ criticality at vanishing and finite net-baryon densities~\cite{Friman:2011pf, Stephanov:2011pb, Karsch:2019mbv, Braun-Munzinger:2020jbk, Braun-Munzinger:2016yjz}. Non-monotonic behavior is expected for various ratios of the cumulants of the net-baryon number. The results of the Beam Energy Scan at the Relativistic Heavy Ion Collider, which covered $\sqrt{s_{\rm NN}}=3-200~$GeV, have shown indications of a non-monotonic behavior in cumulants of the net-proton multiplicity distributions~\cite{STAR:2010mib, STAR:2013gus, Luo:2015ewa, Luo:2017faz, STAR:2022vlo, STAR:2020tga}. However, no definitive consensus has been reached regarding the existence and location of the QCD critical point. More data and higher statistics at low collision energies are needed to draw firm conclusions. 

It has been argued that at large baryon chemical potentials, the chemical freeze-out curve should approach the critical point of the nuclear liquid-gas phase transition~\cite{Floerchinger:2012xd}. The HADES Collaboration at GSI will probe fluctuations at low energies $0.2-1.0A~$GeV, with the expectation of exploring the nuclear liquid-gas phase transition region~\cite{Bluhm:2020mpc}. The properties of nuclear matter and the liquid-gas phase transition have been studied in the literature (see, e.g.,~\cite{Chomaz:2003dz, Pochodzalla:1995xy, INDRA:2001crl, DAgostino:1999nlf, Srivastava:2002xx, ISiS:2001kgh, He:2022yrk, Shao:2020lsv, Deng:2022hjj, Hempel:2013tfa, Koch:2023oez, Marczenko:2023ohi, Vovchenko:2016rkn, Vovchenko:2015pya}).

Due to experimental limitations, only charged particles created in a heavy-ion collision have a fair chance of being measured by the detector. The net-proton number fluctuations are assumed to reliably reflect the net-baryon number fluctuations because proton is the lightest and most abundant baryon created in heavy-ion collisions (HICs). This simplification works for susceptibilities calculated in an ideal gas formulated in a grand-canonical ensemble~\cite{Fukushima:2014lfa}. However, off-diagonal components of susceptibilities can be induced by the presence of interactions \cite{Koch:2023oez}. In HICs, additional effects, such as exact conservation of the baryon number~\cite{Bzdak:2012an, Braun-Munzinger:2020jbk, Braun-Munzinger:2023gsd}, can cause differences between the net-proton and net-baryon cumulants~\cite{Vovchenko:2020kwg}. The baryon cumulants can be reconstructed from the proton cumulants in the case of full randomization of isospin under the assumption that the quantum correlations arising from Fermi statistics are negligible~\cite{Kitazawa:2012at}. However, this is not the case at low energies, where the proton distribution is dominated by interactions between hadrons~\cite{STAR:2022vlo}, the net-proton number fluctuations could cease to be a good proxy for the net-baryon number fluctuations. Qualitative differences in the net-baryon number density fluctuations and individual baryonic chiral partners of opposite parity were examined near the nuclear liquid-gas and chiral phase transitions~\cite{Marczenko:2023ohi, Koch:2023oez, Marczenko:2024jzn}. However, the relations and differences between net-baryon and net-proton number fluctuations have not yet been explored enough in theoretical models near the liquid-gas phase transition.

In this work, we apply the parity doublet model~\cite{Kunihiro:1991qu, Jido:1999hd, Jido:2001nt} to calculate the generalized susceptibilities of the net-baryon number density and the factorial cumulants of the baryon number. Specifically, we focus on the fluctuations of protons and neutrons, as well as the correlations among them. We calculate susceptibilities and factorial cumulants up to the fourth order and examine their qualitative behavior near the nuclear liquid-gas phase transition.

\section{Fluctuations near the nuclear liquid-gas phase transition}
\label{sec:fluct}

\begin{figure}[t]
    \centering
    \includegraphics[width=\columnwidth]{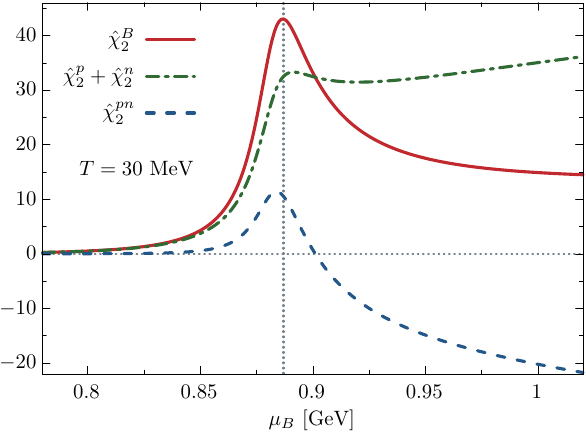}
    \caption{Net-baryon number susceptibility at $T=30~\rm MeV$ in the vicinity of the liquid-gas phase transition. The grey, dotted vertical line marks the crossover transition obtained from the maximum of $\hat \chi_2^B$.}
    \label{fig:x2_t30}
\end{figure}

In this work, we model the liquid-gas phase transition with the hadronic parity doublet model in the mean-field approximation, following~\cite{Koch:2023oez}. We consider a system with $N_f=2$; hence, the groundstate nucleons and their chiral partners are relevant for this study. The hadronic degrees of freedom are coupled to the chiral fields $(\sigma, \boldsymbol\pi)$, and the isosinglet vector field $\omega_\mu$. The thermodynamic potential of the parity doublet model reads
\begin{equation}\label{eq:thermo_full}
    \Omega = \Omega_p + \Omega_n + \Omega_{p^\star} + \Omega_{n^\star} + V_\sigma + V_\omega \textrm,
\end{equation}
where $\Omega_x$'s are the kinetic parts of the thermodynamic potential for positive-parity nucleons, protons ($p$) and neutrons ($n)$ identified with $N(939)$, and their negative-parity chiral partners, $p^\star$ and $n^\star$, identified with $N^\star(1535)$~\cite{ParticleDataGroup:2024cfk}. The potentials $V_{\sigma/\omega}$ are introduced for the scalar ($\sigma$) and vector ($\omega$) mean fields. In-medium profiles of the mean fields are obtained by extremizing the thermodynamic potential in the direction of the mean fields, leading to the gap equations. The set of parameters used in this work and a detailed model description can be found in~\cite{Koch:2023oez}.

At low temperatures, the model predicts sequential first-order nuclear liquid-gas and chiral phase transitions with critical points located at $T_c=16~\rm MeV$, $\mu_B = 909~\rm MeV$ ($n_B = 0.053~ {\rm fm^{-3}} = 0.33~n_{\rm sat}$) and $T_c = 7~\rm MeV$, $\mu_B= 1526~\rm MeV$ ($n_B = 1.25 {\rm~fm^{-3}} = 7.82~n_0$), respectively. Here, we pay attention to the fluctuations near the critical point of the liquid-gas phase transition. Thus, we restrict our considerations to low temperatures and small densities. The vacuum mass of the negative-parity nucleons equals $1.5~$GeV~\cite{ParticleDataGroup:2024cfk}. Effectively, the negative-parity nucleons are thermally suppressed near the liquid-gas phase transition, and the system is predominantly composed of positive-parity nucleons~\cite{Koch:2023oez, Marczenko:2023ohi}. Therefore, in this region, the thermodynamic potential can be reliably restricted to
\begin{equation}\label{eq:thermo_lg}
    \Omega = \Omega_p + \Omega_n + V_\sigma + V_\omega \textrm.
\end{equation}
Consequently, the net-baryon number density is
\begin{equation}
    n_B = -\frac{\dd \Omega}{\dd \mu_B}\Bigg|_T = n_p + n_n \textrm, 
\end{equation}
where $n_{p}~(n_n)$ are the net-proton (net-neutron) number densities. Furthermore, we assume that the system is isospin symmetric so that protons and neutrons are equally populated; hence, $n_p = n_n = 1/2~n_B$.

\begin{figure}[t]
    \centering
    \includegraphics[width=\columnwidth]{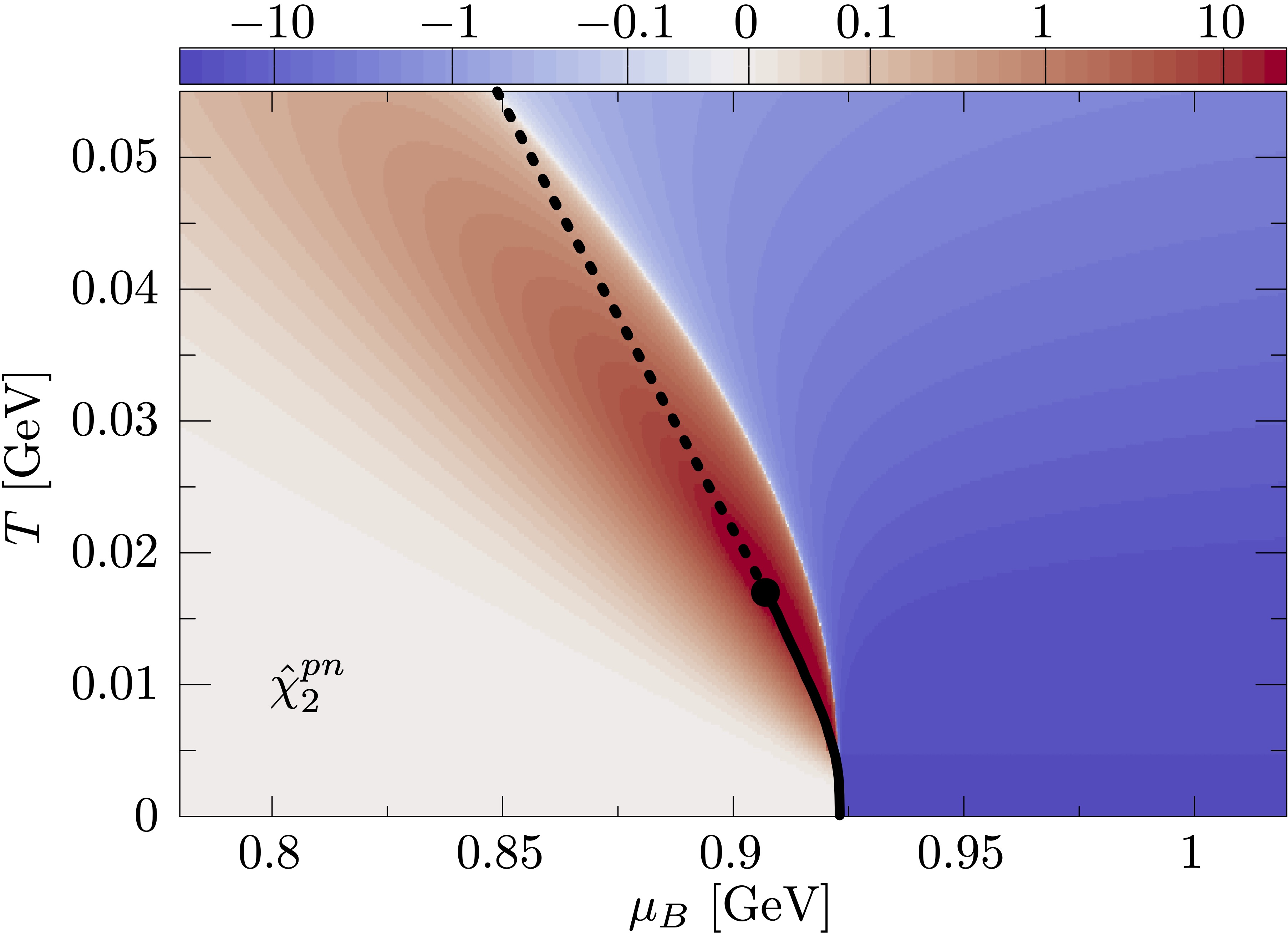}
    \caption{The proton-neutron correlator $\hat\chi_2^{pn}$. The black, solid line marks the first-order liquid-gas phase transition ending with a circle that indicates the critical point. The black, dotted line indicates the crossover transition obtained from the maximum of $\hat\chi_2^B$.}
    \label{fig:heatmap_corr}
\end{figure}

\begin{figure*}[t]
    \centering
    \includegraphics[width=\columnwidth]{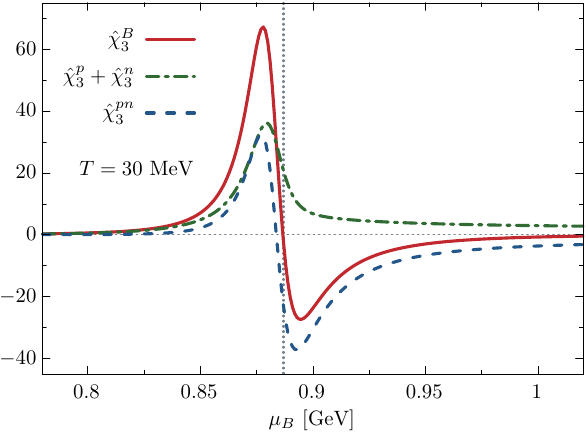}\;
    \includegraphics[width=\columnwidth]{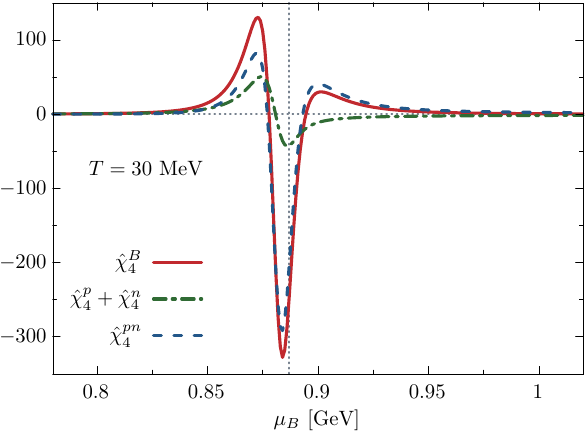}
    \caption{Third-order (left panel) and fourth-order (right panel) susceptibilities of the net-baryon number at $T=30~\rm MeV$ in the vicinity of the liquid-gas phase transition. The grey, dotted vertical line marks the crossover transition obtained from the maximum of $\hat \chi_2^B$.}
    \label{fig:x3x4_t30}
\end{figure*}

In the grand canonical ensemble, the generalized susceptibilities of the net-baryon number, $\chi_k^B$, are defined as derivatives with respect to the baryon chemical potential,
\begin{equation}
    \hat \chi_k^B = -\frac{\dd^k \hat\Omega}{\dd \hat\mu_B^k}\Bigg|_T \textrm,
\end{equation}
where the hat symbol indicates that a quantity is temperature-normalized. To be able to connect the susceptibilities of the net-proton and net-neutron numbers to the net-baryon
susceptibilities, one needs to rewrite the mean-field thermodynamic potential in Eq.~\eqref{eq:thermo_lg} in terms of newly defined independent chemical potentials, $\mu_p$ and $\mu_n$, for protons and neutrons as follows~\footnote{Detailed derivation can be found in Ref.~\cite{Koch:2023oez}.}:
\begin{equation}\label{eq:thermo_pm}
\begin{split}
    \Omega &=\Omega_p\left(\mu_p, T, \sigma\left(\mu_p, \mu_n\right), \omega\left(\mu_p, \mu_n\right)\right)\\
    &+ \Omega_n\left(\mu_n, T, \sigma\left(\mu_p, \mu_n\right), \omega\left(\mu_p, \mu_n\right)\right) \\
    &+ V_\sigma(\sigma\left(\mu_p, \mu_n\right)) + V_\omega(\omega\left(\mu_p, \mu_n\right))\rm.
\end{split}
\end{equation}
Such a separation into separate chemical potentials is possible in the mean-field approximation, which is a single-particle theory (see detailed discussion in~\cite{Garcia:2018iib}). This allows us to take derivatives directly with respect to $\mu_{p/n}$. 

The second-order susceptibility of the net-baryon number can be written as follows
\begin{equation}\label{eq:x2_pn}
    \hat\chi_{2}^B = \hat\chi_2^{p}  + \hat\chi_2^{n} + \hat\chi_2^{pn}\textrm,
\end{equation}
where $\hat\chi_2^{p}$ and $\hat\chi_2^{n}$ are the susceptibilities of the net-proton and net-neutron numbers,
\begin{equation}\label{eq:susc_2}
    \hat \chi_2^{p/n} = -\frac{\dd^2 \hat \Omega}{\dd \hat\mu_{p/n}^2}\Bigg|_{T}\textrm,
\end{equation}
respectively. We note that $\hat\chi_2^{p} = \hat \chi_2^{n}$, due to isospin symmetry assumed in this work. The last term, $\hat\chi_2^{pn}$, is the correlation between the net-proton and net-neutron numbers,
\begin{equation}\label{eq:corr_2}
    \hat \chi_2^{pn} = -2\frac{\dd^2 \hat \Omega}{\dd \hat\mu_p \,\dd \hat\mu_n}\Bigg|_{T}\textrm.
\end{equation}

\begin{figure*}[t]
    \centering
    \includegraphics[width=0.7\linewidth]{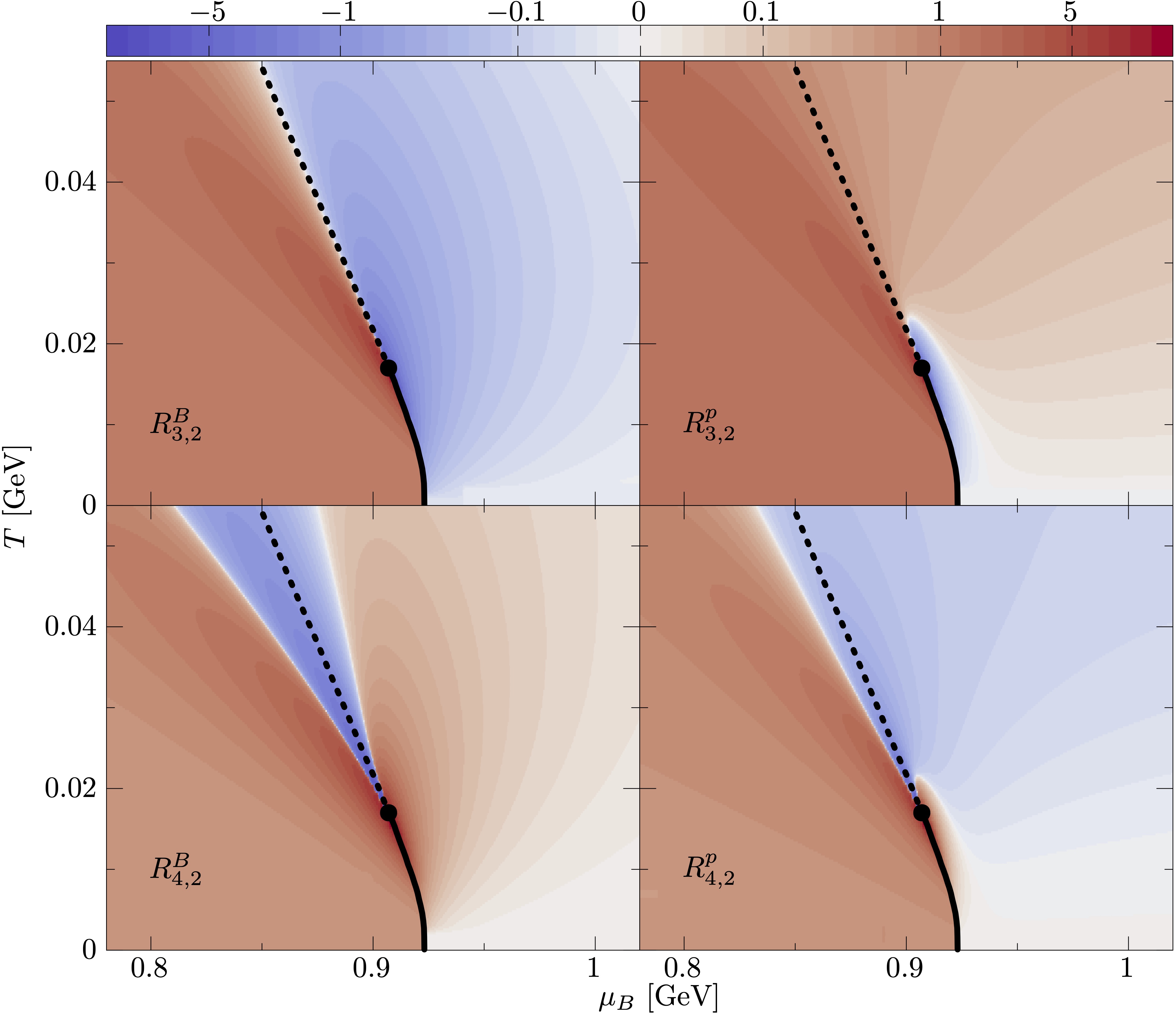}
    \caption{Skewness $R_{3,2}^\alpha$ (top panel) and kurtosis $R_{4,2}^\alpha$ (bottom panel) for the net-baryon (left panel) and net-proton number (right panel) densities. The black, solid line marks the first-order liquid-gas phase transition ending with a circle that indicates the critical point. The black, dotted line indicates the crossover transition obtained from the maximum of $\hat\chi_2^B$.}
    \label{fig:heatmap_ratios}
\end{figure*}

Fig.~\ref{fig:x2_t30} shows the susceptibility of the net-baryon number density, $\hat\chi_2^B$, at $T=30~\rm MeV$ in the vicinity of the liquid-gas phase transition. It features a clear maximum as it passes through the crossover transition and decreases afterward. Initially, at small baryon chemical potential, the susceptibility $\hat\chi_2^{p}$ also increases and features a maximum in the vicinity of the crossover transition, followed by a shallow minimum, after which it continues to grow. Notably, as the crossover transition is approached, $\hat\chi_2^p$ starts to deviate from $\hat\chi_2^B$ and has an opposite trend after the transition. This already indicates that the correlator $\hat \chi_2^{pn}$ is non-vanishing and has a non-trivial structure [{\it cf.}~Eq.~\eqref{eq:x2_pn}]. We find that at small chemical potentials, $\hat\chi_2^{pn} \simeq 0$ and $\hat\chi_2^B \simeq 2\hat\chi_2^{p}$. However, as the crossover transition is approached, the correlator increases, a peak is observed in the vicinity of the transition, and it continues to decrease, eventually becoming negative. This causes an enhancement of $\hat\chi_2^p$ compared to $\hat\chi_2^B$ after the transition. It should be noted that experimentally measurable fluctuations of the proton number are identified as proxies of the baryon number fluctuations. From Eq.~\eqref{eq:susc_2}, it is clear that such an assumption requires a vanishing $\hat\chi_2^{pn}$ correlator. Our results indicate that this is not the case, especially in the vicinity of a CP. In fact, susceptibilities such as the up-down quark correlator $\hat\chi_2^{ud}$ are not vanishing at finite temperature in LQCD simulations~\cite{Gavai:2005yk} and model calculations~\cite{Mukherjee:2006hq, Cristoforetti:2010sn}. 

Using $\mu_p = 2\mu_u + \mu_d$ and $\mu_n = \mu_u + 2\mu_d$, the correlator $\hat\chi_2^{ud}$ can be expressed as
\begin{equation}
    \hat \chi_2^{ud} = 4\hat \chi_2^{p} + 5 \hat \chi_2^{pn} \textrm,
\end{equation}
where we have assumed that $\hat \chi_2^p = \hat \chi_2^n$. Even in the absence of correlations, $\hat \chi_2^{ud} \sim \hat \chi_2^{p}$ is non-vanishing due to the quark substructure of nucleons~\cite{Bellwied:2015lba}. Any qualitative difference between $\hat \chi_2^{ud}$ and $\hat \chi_2^{p}$ amounts to the non-trivial proton-neutron correlations. Here, they appear due to the presence of interactions mediated by the mean fields \mbox{[{\it cf.}~\eqref{eq:thermo_pm}]}. We point out that in a more general case mesons contribute to $\hat \chi_{2}^{ud}$ as well with negative sign~\cite{Gavai:2005yk} at finite temperature.

In Fig.~\ref{fig:heatmap_corr}, we show the behavior of the correlator $\hat\chi_2^{pn}$ in the $\mu_B-T$ plane. At any temperature shown, $\hat\chi_2^{pn} \rightarrow 0$ as $\mu_B$ approaches zero, indicating that the system behaves as a free gas of nucleons. In the vicinity of a CP, the correlator increases substantially, develops a maximum, and turns negative at larger chemical potentials. Due to the non-trivial structure of the $\hat\chi_2^{pn}$ correlator near the CP, a common identification of the net-proton number fluctuations as half of the net-baryon number fluctuations is violated. This result is obtained in the isospin-symmetric case, i.e., $n_B = 2~n_{p/n}$, with complete randomization of the isospin space.

Higher-order fluctuations are expected to reveal potential critical structures related to the existence of a CP. The $k^{\rm th}$ order susceptibilities and correlations are defined as follows
\begin{equation}\label{eq:high_susc}
    \hat \chi_k^{p/n} = -\frac{\dd^k \hat \Omega}{\dd \hat\mu_{p/n}^k}\Bigg|_{T}
\end{equation}
and
\begin{equation}\label{eq:high_corr}
    \hat \chi_k^{pn} = -\sum_{j=1}^{k-1}\binom{k}{j}\frac{\dd^k \hat \Omega}{\dd \hat\mu_{p}^j \,\dd \hat\mu_n^{k-j}}\Bigg|_{T}\textrm,
\end{equation}
respectively. The above definitions are direct generalizations of Eqs.~\eqref{eq:susc_2} and~\eqref{eq:corr_2}, and
\begin{equation}\label{eq:susc_k}
    \hat \chi_k^B = \hat \chi_k^p + \hat \chi_k^n + \hat \chi_k^{pn} \textrm.
\end{equation}
In our case, $\hat \chi_k^p = \hat \chi_k^n$ due to isospin symmetry. We note that each term in Eq.~\eqref{eq:high_corr} could be studied in principle individually, revealing more complex structures of the correlator. We plan to elaborate on this elsewhere.

Fig.~\ref{fig:x3x4_t30} shows the third and fourth-order susceptibilities at $T=30~$MeV. Both susceptibilities of the net-baryon number density ($\hat\chi_{3}^B$ and $\hat \chi_4^B$) exhibit distinctive behavior and show strong sensitivity to dynamical effects related to the liquid-gas phase transition. Interestingly, we observe that susceptibilities $\hat\chi_{3,4}^p$ do not qualitatively reflect $\hat\chi_{3,4}^B$. The third-order susceptibility $\hat\chi_{3}^p$ is positive and shows a maximum in the vicinity of the crossover transition. The fourth-order susceptibility $\hat\chi_{4}^p$ features a peak-dip structure. However, it does not turn positive after the crossover transition, compared to $\hat\chi_{4}^B$. In contrast to this, the correlators $\hat \chi_{3}^{pn}$ and $\hat \chi_{4}^{pn}$ have structures that qualitatively resemble those of $\hat\chi_{3}^B$ and $\hat\chi_{4}^B$, respectively.

\begin{figure}[t]
    \centering
    \includegraphics[width=1\linewidth]{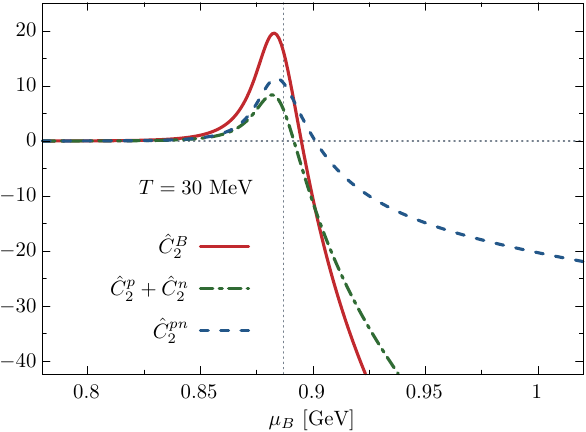}
    \caption{Second-order factorial cumulants, $\hat C^\alpha_2$, at $T=30~\rm MeV$ in the vicinity of the liquid-gas phase transition. The grey, dotted vertical line marks the crossover transition obtained from the maximum of $\hat \chi_2^B$.}
    \label{fig:C2}
\end{figure}

Non-monotonic behavior is also expected for various ratios of the higher-order fluctuations of the net-baryon number. Ratios of event-by-event cumulants and correlations are considered in experiments because they cancel out the leading-order volume dependence and its fluctuations, which are unknown in heavy-ion collisions. We define ratios of the susceptibilities as follows
\begin{equation}\label{eq:ratio_def}
    R_{n,k}^{\alpha} = \frac{\hat\chi^{\alpha}_n}{\hat\chi_k^{\alpha}}\textrm,
\end{equation}
where $\alpha=B,p,n$ denotes the net number of baryons, protons, and neutrons, respectively. Particularly useful are skewness $R_{3,2}^\alpha$ and kurtosis $R_{4,2}^\alpha$. They are depicted in Fig.~\ref{fig:heatmap_ratios} in the $\mu_B-T$ plane. At small chemical potentials, these ratios are equal to unity due to the expected Skellam probability distribution, namely $R^\alpha_{3,2} = \tanh{\left(\mu_B/T\right)}$ and $R^\alpha_{4,2} = 1$~\cite{BraunMunzinger:2003zd}. Close to the transition, the ratios for the net-baryon number feature structures that are expected at the phase boundary. The low-density phase corresponds to $R_{3,2}^B>0$, while the high-density phase corresponds to $R_{3,2}^B<0$. The crossover line corresponds to $R_{3,2}^B=0$. The crossover region is also characterized by negative kurtosis $R_{4,2}^B$. However, around this negative region, it attains large positive values. These critical structures are qualitatively expected from model-independent universality arguments~\cite{Stephanov:2008qz, Stephanov:2011pb}. The ratios for the net-proton number density ($R_{3,2}^p$ and $R_{4,2}^p$) are qualitatively different. In general, we observe that the critical region shrinks considerably when the net-proton number density fluctuations are considered as opposed to the net-baryon number density fluctuations. For example, $R_{3,2}^p$ is positive at $T \gtrsim 20~$MeV, and it turns negative after the transition in the close vicinity of the CP. This is in contrast to $R_{3,2}^B$, which already shows this behavior at much higher temperatures.

\begin{figure*}[t]
    \centering
    \includegraphics[width=0.48\linewidth]{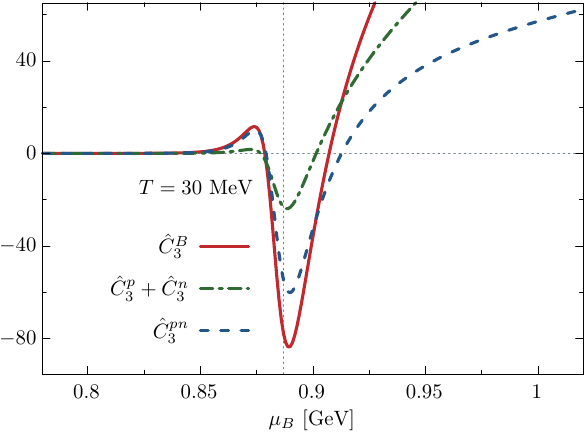}\;
    \includegraphics[width=0.48\linewidth]{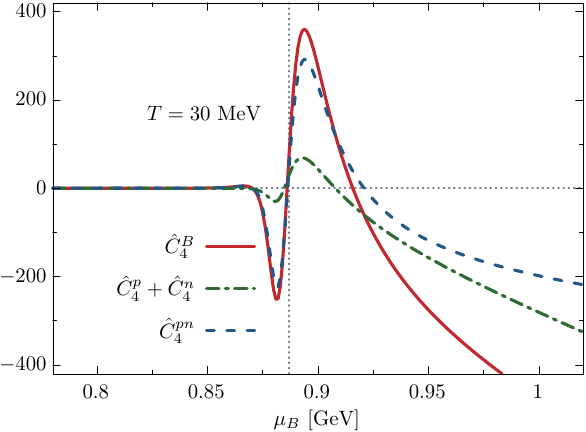}
    \caption{Third-order (left panel) and fourth-order (right panel) factorial cumulants of the baryon and proton numbers at $T=30~\rm MeV$ in the vicinity of the liquid-gas phase transition. The grey, dotted vertical line marks the crossover transition obtained from the maximum of $\hat \chi_2^B$.}
    \label{fig:C3_4_30}
\end{figure*}

In a two-component system of protons and neutrons, the fluctuations of conserved charges, net-baryon ($B$) and net-charge ($Q$) numbers can be identified with specific mixtures of $\hat\chi_2^{p/n}$. The relations for the net-baryon number susceptibilities are given in Eqs.~\eqref{eq:susc_2} and~\eqref{eq:susc_k}. For the second-order charge fluctuations and baryon-charge correlator one gets
\begin{align}\label{eq:xbb2}
    \hat \chi_2^Q    &=-\frac{\dd^2 \hat \Omega}{\dd \hat\mu_Q^2}\Bigg|_T =  \hat \chi_2^p  \textrm,\\
    \hat \chi_2^{BQ} &= -\frac{\dd^2 \hat \Omega}{\dd \hat\mu_B \dd \hat\mu_Q}\Bigg|_T = \hat \chi_2^p + \frac{1}{2} \hat \chi_2^{pn} = \frac{1}{2} \hat \chi_2^B\textrm,\label{eq:xbq}
\end{align}
respectively. We note that the last equality in Eq.~\eqref{eq:xbq} is valid due to isospin symmetry. Similarly for the third- and fourth-order susceptibilities
\begin{align}\label{eq:xbq34}
    \hat \chi_3^Q &=- \frac{\dd^3 \hat \Omega}{\dd \hat\mu_Q^3}\Bigg|_T = \hat \chi_3^p\\
    \hat \chi_4^Q &=- \frac{\dd^4 \hat \Omega}{\dd \hat\mu_Q^4}\Bigg|_T = \hat \chi_4^p
\end{align}
We have verified the above relations also numerically. Various baryon-charge correlators at higher orders are given by a mixture of correlation functions. We elaborate on these in our future work. 

In the vicinity of the liquid-gas phase transition, the nuclear medium is predominantly comprised of protons and neutrons. Therefore, one expects that, $\hat \chi_k^Q \simeq \hat \chi_k^p$, as derived in Eqs.~\eqref{eq:xbb2} and~\eqref{eq:xbq34}. Moreover, the results obtained in our work qualitatively agree with other approaches to nuclear liquid-gas phase transition based on the quantum van der Waals interactions~\cite{Poberezhnyuk:2019pxs}. However, we note that the above does not need to be the case at higher temperature and/or baryon chemical potential, where contributions from the other baryons and their resonances are to be expected. 

We note that the nuclear symmetry energy controls the change of the binding energy of nuclear matter as the proton-to-neutron ratio changes. Thus, higher values of the symmetry energy should suppress the individual proton and neutron fluctuations and enhance the correlations among them. However, in the present work, we have not considered the effect of the symmetry energy. Phenomenologically, to fit the physical value for the symmetry energy, the $\rho$ meson coupling to nucleons has to be added to the model Lagrangian. We leave this interesting and important issue for our future work.

\section{Factorial Cumulants}

Baryon number cumulants can be reconstructed from proton cumulants measured experimentally under the assumption of total isospin randomization in the final state of a heavy-ion collision~\cite{Kitazawa:2012at}. In this context, useful are factorial cumulants ($\hat C_n$) as they eliminate self-correlations. Then, the proton number factorial cumulants $\hat C_n^p$ are directly proportional to baryon number factorial cumulants ($\hat C^B_n$),
\begin{equation}\label{eq:kitazawa}
    \hat C_n^B = 2^n \hat C_n^p \textrm.
\end{equation}
However, it should be noted that this proportionality is only valid under the assumption that the correlations arising from quantum Fermi statistics are negligible. This is justified at high energies $\sqrt{s_{NN}} \gtrsim 10~$GeV~\cite{Kitazawa:2012at}, which corresponds to the freeze-out temperature $T \gtrsim 150~$MeV~\cite{Andronic:2017pug}. This is not necessarily the case at low energies, where the proton distribution is dominated by interactions between hadrons~\cite{STAR:2022vlo}, and especially in the vicinity of the nuclear liquid-gas phase transition, where Fermi statistics effects are certainly non-negligible.

Factorial cumulants can be expressed as linear combinations of the ordinary susceptibilities~\cite{Bialas:2007ed, Bzdak:2016sxg, Friman:2022wuc}. The first four factorial cumulants read:
\begin{align}
    \hat C^\alpha_1 &= \hat \chi_1^\alpha \textrm, \label{eq:fac_cum_def1}\\
    \hat C^\alpha_2 &= \hat \chi_2^\alpha - \hat\chi_1^\alpha \textrm, \\
    \hat C^\alpha_3 &= \hat \chi_3^\alpha - 3\hat\chi_2^\alpha + 2\hat\chi_1^\alpha \textrm, \\
    \hat C^\alpha_4 &= \hat \chi_4^\alpha - 6\hat\chi_3^\alpha + 11\hat\chi_2^\alpha - 6 \hat\chi_1^\alpha \textrm,\label{eq:fac_cum_def4}
\end{align}
where $\alpha = B, p, n$. Typically, factorial cumulants, $\hat C_n$, are defined as linear combinations of ordinary cumulants, $\hat \kappa_n$, which are related to the susceptibilities through $\kappa_n = VT^3\hat \chi_n$. Therefore, the definition of factorial cumulants in Eqs.~\eqref{eq:fac_cum_def1}-\eqref{eq:fac_cum_def4} differ by the factor $VT^3$ from usual definition through ordinary cumulants. Nevertheless, our definition does not lead to qualitative differences and the prefactors cancel out when ratios of cumulants are considered. We note that the correlation functions $\hat C^\alpha_k$ are defined for particles and not the net-particle numbers. 
However, in the vicinity of the liquid-gas phase transition, the contribution of antiparticles is well suppressed and we can stick to the relations above. Nevertheless, a direct relation between the factorial cumulants and susceptibilities for net-particle numbers can also be derived~\cite{Bzdak:2016sxg, Friman:2022wuc}. In the isospin-symmetric system, $\hat C_k^p = \hat C_k^n$.

\begin{figure*}[t]
    \centering
    \includegraphics[width=0.7\linewidth]{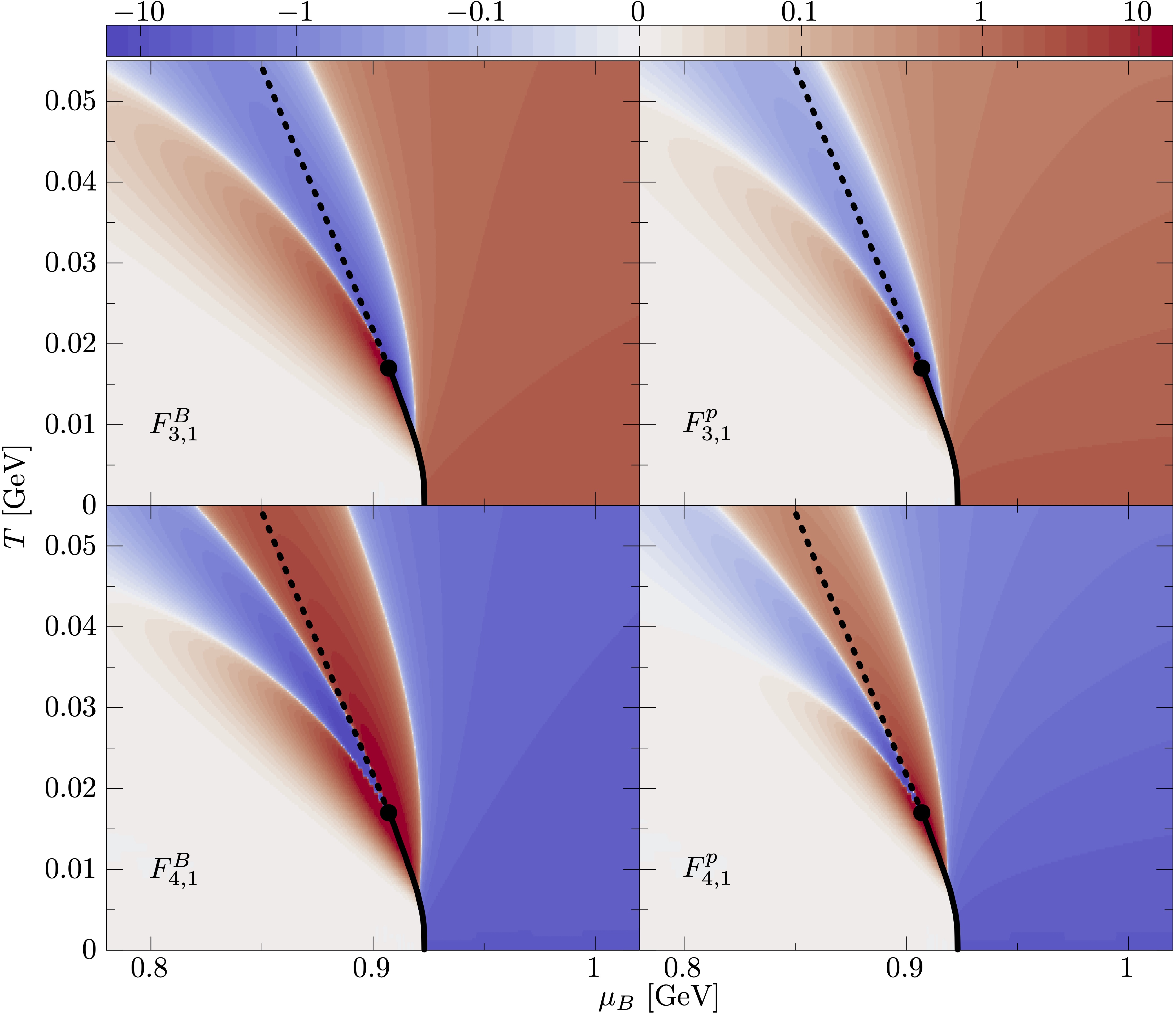}
    \caption{Ratio $F_{3,1}^\alpha$ (top panel) and $F_{4,1}^\alpha$ (bottom panel) for factorial cumulants of baryons (left panel) and protons (right panel). The black, solid line marks the first-order liquid-gas phase transition ending with a circle that indicates the critical point. The black, dotted line indicates the crossover transition obtained from the maximum of $\hat\chi_2^B$.}
    \label{fig:heatmap_ratios_C}
\end{figure*}

The second-order factorial cumulant for the baryon number at $T=30~$MeV is depicted in Fig.~\ref{fig:C2}. It develops a peak around the remnant of the liquid-gas phase transition and turns negative, driven by the repulsive interactions. Notably, the factorial cumulant for protons, $\hat C_2^p$, resembles qualitatively the same structure. Although $\hat C_2^B \simeq C_2^p$, we find that their difference has a non-trivial structure as well. The difference between baryon number factorial cumulants and the factorial cumulants for individual species, which we denote as $\hat C_k^{pn}$, are given as follows:
\begin{align}
    \hat C_1^{pn} \equiv \hat C_1^B - \sum_{\alpha=p,n}\hat C_1^\alpha&= 0 \textrm,\label{eq:frac_diff1}\\
    \hat C_2^{pn} \equiv \hat C_2^B -  \sum_{\alpha=p,n}\hat C_2^\alpha&= \hat \chi_2^{pn}  \textrm,\\
    \hat C_3^{pn} \equiv \hat C_3^B -  \sum_{\alpha=p,n}\hat C_3^\alpha&= \hat \chi_3^{pn} - 3\hat \chi_2^{pn}  \textrm,\\
    \hat C_4^{pn} \equiv \hat C_4^B -  \sum_{\alpha=p,n}\hat C_4^\alpha&= \hat \chi_4^{pn} - 6\hat \chi_3^{pn} + 11\hat \chi_2^{pn} \label{eq:frac_diff4}\textrm.
\end{align}
From the above equations, one sees that the difference between $\hat C_2^B$ and $\hat C_2^p$ amounts to the proton-neutron correlator, $\hat \chi_2^{pn}$. Departure of $\hat C_k^{pn}$ from zero, signals physics beyond the uncorrelated gas of nucleons, which is expected in the vicinity of the liquid-gas phase transition. We note that the relations in Eqs.~\eqref{eq:frac_diff1}-\eqref{eq:frac_diff4} do not require additional assumptions and are expected to work for any temperature and chemical potential. In general $\hat C_k^{pn} \neq 0$ breaks the scaling in Eq.~\eqref{eq:kitazawa}. Nevertheless, it is possible to retain it, if $\hat C_k^{pn} = \left(2^k -2 \right)\hat{C}_k^p$. However, we do not find this to be the case in general and $\hat C_k^{pn}$ is not proportional to $\hat C_k^p$ [{\it cf.}~Figs.~\ref{fig:C2} and~\ref{fig:C3_4_30}].

In Fig.~\ref{fig:C3_4_30}, we show the third- and fourth-order factorial cumulants. The proton number factorial cumulants behave qualitatively similar to the baryon number factorial cumulants. Interestingly, the difference between them show the same qualitative features as $\hat C_k^B$ and $\hat C_k^p$. This is in clear contrast to the susceptibilities [{\it cf.} Figs.~\ref{fig:x2_t30} and~\ref{fig:x3x4_t30}].

Similarly to the susceptibilities, we define ratios of factorial cumulants,
\begin{equation}
    F_{n, k}^\alpha = \frac{\hat C^\alpha_n}{\hat C^\alpha_k}
\end{equation}
for $\alpha = B, p, n$. In Fig.~\ref{fig:heatmap_ratios_C}, we show the ratios of the factorial cumulants $F_{3,1}^\alpha$ and $F_{4,1}^\alpha$. The structure of factorial-cumulant ratios for protons and baryons remains qualitatively the same for $F_{3,1}^\alpha$ and $F_{4,1}^\alpha$. Nevertheless, their differences can be shown to be directly proportional to the mixture of correlation functions,
\begin{align}
    F_{2, 1}^B - F_{2, 1}^p &= \frac{\hat C_2^{pn}}{\hat C_1^B} \textrm,\\
    F_{3, 1}^B - F_{3, 1}^p &= \frac{\hat C_3^{pn}}{\hat C_1^B} \textrm,\\
    F_{4, 1}^B - F_{4, 1}^p &= \frac{\hat C_4^{pn}}{\hat C_1^B} \textrm,
\end{align}
where we have assumed isospin symmetry. The differences between the ratios are directly proportional to the differences derived in Eqs.~\eqref{eq:frac_diff1}-\eqref{eq:frac_diff4}.

Factorial cumulants eliminate self-correlations that arise from trivial contributions in regular susceptibilities (or cumulants). This allows us to identify the genuine correlations in the system. We find that the presence of interactions, i.e., $\hat \chi_k^{pn} \neq 0$, modifies the scaling $\hat C_k^B = 2^k \hat C_k^p$, which is expected since the difference between $\hat C_k^B$and  $\hat C_k^p$ is directly proportional to the linear combination of correlations up to $k$-th order. We find that although the baryon and proton cumulants have qualitatively the same structure in the vicinity of the liquid-gas phase transition, their difference is also non-trivial, showing similar features.

\section{Summary}
\label{sec:summary}

We investigated the net-baryon number density fluctuations close to the nuclear liquid-gas phase transition in the isospin-symmetric matter. To this end, we used the parity doublet model in the mean-field approximation to account for the critical behavior. In particular, we have studied the susceptibilities of the net-proton and net-neutron numbers, as well as their correlations up to the fourth order. These quantities show nonmonotonic structures at the phase boundary and are one of the main tools used in experimental searches for the QCD critical point. We discussed the effects of the correlations between protons and neutrons and found that they are non-trivial. Consequently, the net-proton number fluctuations have a qualitatively different structure than the net-baryon number fluctuations. 

We have also studied the factorial cumulants. We find that the qualitative structure of the baryon-number factorial cumulant is preserved when the proton-number factorial cumulant is considered. Furthermore, the difference between baryon- and proton-number factorial cumulants is directly proportional to a linear combination of the proton-neutron correlation functions. We demonstrated that although it is nontrivial, it has a qualitatively similar structure to the factorial cumulants themselves. We expect these results to hold in the vicinity of the liquid-gas phase transition.

Our results emphasize the importance of the interactions between baryons for higher-order fluctuations. In particular, they bring significant and nontrivial differences in the critical behavior of the net-proton and net-baryon number density fluctuations close to the critical endpoint of the liquid-gas phase transition. We hope they could be useful in searching for critical points in the QCD phase diagram and interpreting the critical properties of the matter produced in heavy-ion collisions. 

Finally, we note that the properties of the susceptibilities were obtained with the parity doublet model in the mean-field approximation. Nevertheless, the critical properties of the proton-neutron correlations near the liquid-gas phase transition appear because of the presence of interactions mediated via mean-field potentials. Therefore, they are expected to be quite generic. Thus, the results presented in this work are robust on a qualitative level.

\section*{Acknowledgments}
This work was supported by the program Excellence Initiative–Research University of the University of Wroc\l{}aw of the Ministry of Education and Science (M.M.). This work is partly funded by the Polish National Science Centre (NCN) under OPUS Grant No. 2022/45/B/ST2/01527 (C.S. and K.R.). K.R. also acknowledges the support of the Polish Ministry of Science and Higher Education. The work of C.S. was supported in part by the World Premier International Research Center Initiative (WPI) under MEXT, Japan. We acknowledge interesting discussions with Volker Koch and Larry McLerran. We also thank the anonymous referee for their constructive comments and suggestion to extend the study to factorial cumulants.

\bibliography{biblio}

\end{document}